\documentclass{aa}  

\usepackage{graphicx}
\usepackage{txfonts}
\usepackage{subcaption}         
\usepackage{lscape}             
\usepackage{placeins}           
\usepackage{lipsum}

\usepackage[encapsulated]{CJK}
\usepackage{ucs}
\usepackage[utf8]{inputenc}
\usepackage[colorlinks=true, allcolors=blue]{hyperref}

\newcommand{\tctext}[1]{\begin{CJK}{UTF8}{bkai}#1\end{CJK}}
\newcommand{\tktext}[1]{\begin{CJK}{UTF8}{mj}#1\end{CJK}}
\newcommand{\amnh}{Department of Astrophysics, American Museum of Natural History, 200 Central Park West, New York, NY 10024, USA}

\begin{document}

   \title{Conditions for planetesimal formation via the streaming instability persist under turbulence driven by magnetorotational instability}
   \titlerunning{Streaming instability and MRI turbulence}
   
   \author{Linn E. J.~Eriksson\inst{1}
          \and Ziyan~Xu\inst{2}
          \and Jeonghoon~Lim (\tktext{임정훈})\inst{3}
          \and Chao-Chin~Yang (\tctext{楊朝欽})\inst{4}
          \and Pinghui~Huang\inst{5}
          \and Mordecai-Mark Mac~Low\inst{1}
          }
          
    \institute{\amnh\\
              \email{leriksson@amnh.org}
              \and Center for Star and Planet Formation, GLOBE Institute, University of Copenhagen, Øster Voldgade 5–7, 1350 Copenhagen, Denmark
              \and Department of Physics and Astronomy, Iowa State University, Ames, IA 50010, USA
              \and Department of Physics and Astronomy, The University of Alabama, PO Box 870324, Tuscaloosa, AL 35487-0324, USA
              \and CAS Key Laboratory of Planetary Sciences, Purple Mountain Observatory, Chinese Academy of Sciences, Nanjing 210023, People’s Republic of China
             }

   \date{Received Month day, year}

  \abstract
  {Strong dust clumping by streaming instability (SI) is the leading proposed mechanism for forming planetesimals, the building blocks of terrestrial planets and giant-planet cores. The critical dust-to-gas density ratio above which the SI leads to dust concentration strong enough to result in gravitational collapse depends on local dust properties and disk conditions, such as particle Stokes number, pressure gradient, and turbulence. The role of turbulence has recently drawn attention because simulations have shown that even modest levels of istropically forced turbulence can significantly increase the critical dust-to-gas ratio. However, we show that this does not hold for turbulence self-consistently generated by the magnetorotational instability (MRI). We present the first parameter study of the SI in three-dimensional shearing-box simulations including non-ideal magnetohydrodynamics with ambipolar diffusion. Modest turbulence yields a clumping boundary as low as pure hydrodynamical cases, while stronger turbulence does increase the critical dust-to-gas density ratio, though appreciably less than in the models where turbulence is isotropically forced. Particle concentration occurs inside zonal flows, large-scale structures generated by the MRI. Our results suggest that self-consistent, MRI-driven turbulence does not necessarily inhibit planetesimal formation.
 }

   \maketitle

\section{Introduction}\label{sec:introduction}

The first step in the planet formation process is the collisional coagulation of micron-sized dust grains into roughly millimeter-sized pebbles. Continued growth toward planetesimal sizes is hindered by multiple growth barriers, most notably the fragmentation barrier (e.g. \citealt{BlumWurm2008, Guttler2010}) and the radial drift barrier (e.g. \citealt{Weidenschilling1977, Birnstiel_2012}). Highly porous dust aggregates have been proposed as a possible way to circumvent these barriers, but this remains controversial (see discussion in \citealt{Drazkowska2023}). More attention has been paid to planetesimal formation via the direct gravitational collapse of dense pebble clumps. Multiple mechanisms can concentrate solids to planetesimal-forming densities, but currently the most promising one is the streaming instability (SI; \citealt{YoudinGoodman2005,JohansenYoudin2007}). The SI arises from the exchange of angular momentum between gas and solid particles via aerodynamic coupling and leads to the formation of radially narrow particle filaments \citep{YangJohansen2014,Li2018} that can reach densities high enough for gravitational collapse \citep{Johansen2015,Simon2016}. SI simulations have successfully produced planetesimal populations consistent with properties of Solar System minor bodies, including the size distribution of the asteroids \citep{Johansen2015} and cold classical Kuiper Belt objects (KBOs) \citep{Kavelaars2021}, the orbital properties of trans-Neptunian binaries \citep{Nesvorny2019}, and the shapes \citep{Barnes2025} and prevalence of contact binaries \citep{Barnes2026} among the cold classical KBOs.

Identifying the conditions under which the SI leads to planetesimal formation is an active area of research. The growth rate of the linear, unstratified SI depends mainly on the particle-to-gas volume density ratio $\epsilon$ and the Stokes number
\begin{equation}
\mathrm{St} \equiv t_{\rm stop}\Omega_{\rm K},
\end{equation}
also referred to as the dimensionless stopping time, where the stopping time $t_{\rm stop}$ is proportional to the particle size and $\Omega_\mathrm{K}$ is the Keplerian orbital frequency \citep{YoudinGoodman2005}. Early studies of the SI showed that both the linear growth rate and nonlinear clumping are stronger when $\epsilon\gtrsim1$ \citep{YoudinGoodman2005,JohansenYoudin2007_turbulence}. In the nonlinear regime, the clumping criterion for the monodisperse SI has been extensively investigated using 2D axisymmetric, vertically stratified simulations \citep{Carrera2015,Yang2017,LiYoudin2021,Lim2025}. These studies have quantified the boundary above which clumping occurs in terms of $\mathrm{St}$, together with $\epsilon$, the particle-to-gas surface density ratio $Z$, or both. In addition, the clumping boundary depends on the radial pressure gradient, with lower pressure gradients allowing clump formation at lower particle-to-gas densities for a given $\mathrm{St}$ \citep{BaiStone2010b,Sekiya2018}. The most recent of these studies, \citet{Lim2025}, found that the clumping boundary lies at subsolar metallicities for $\mathrm{St}\gtrsim0.01$. 

\defcitealias{Lim2026}{L26}
Extending these results to three dimensions (3D), \citet[][hereafter \citetalias{Lim2026}]{Lim2026}, performed a parameter study of the monodisperse SI. The transition from 2D to 3D simulations results in further reduced (subsolar) $Z_{\rm crit}$ for $\mathrm{St}\gtrsim0.03$, increased to supersolar $Z_{\rm crit}$ for $\mathrm{St}\leq0.02$, and a sharp transition between these two values of $\mathrm{St}$. They derived the following fit to the strong-clumping boundary:
\begin{equation}\label{eq: L25}
\log Z_{\rm crit}(\mathrm{St}) = A(\log\mathrm{St})^2 + B\log\mathrm{St} + C \quad \mathrm{for}\;\, \mathrm{St}>0.025,
\end{equation}
where the fitting parameters are given in Table \ref{table: fit params} and $Z_{\rm crit}$ is the critical dust-to-gas surface density ratio above which the maximum particle density exceeds the Hill density.

\defcitealias{Lim2024}{L24}
All these criteria were derived in the absence of external turbulence; however, observations suggest that disks are at least weakly turbulent (e.g. \citealt{Villenave+22,Pinte+23}). \citet[][hereafter \citetalias{Lim2024}]{Lim2024} performed a parameter study of the monodisperse SI under externally driven (forced) turbulence using 3D stratified simulations that include self-gravity of particles. They obtained the following fit to the clumping boundary for gravitational collapse:
\begin{align}\label{eq: L24}
    \log Z_{\rm crit}(\textrm{St},\alpha_{z,\rm{D}}) = &A'(\log\alpha_{z,\rm{D}})^2 + B'\log\textrm{St}\log\alpha_{z,\rm{D}} + \nonumber\\ &C'\log\textrm{St} + D'\log\alpha_{z,\rm{D}},
\end{align}
where the fitting parameters are given in Table \ref{table: fit params}. The fit is tested over the range $\alpha_{\rm D} \in [10^{-4},10^{-3}]$ and $\textrm{St} \in [10^{-2},10^{-1}]$. The original fit was presented only in terms of the turbulent parameter for the spatial diffusion of turbulent gas, $\alpha_{\rm D}$. We use the fitting parameters for $\alpha_{\rm D}$ and the relationship between $\alpha_{\rm D}$ and the turbulent parameter for vertical diffusion, $\alpha_{z}$, to calculate the fitting parameters for $\alpha_{\rm z}$ as well.

\begin{table}
\caption{Fitting parameters for the SI boundary by \citetalias{Lim2026} (eq. \ref{eq: L25}) and \citetalias{Lim2024} (eq. \ref{eq: L24}).}
\label{table: fit params}
\begin{tabular}{lllll}
\hline\hline
No turb. (\citetalias{Lim2026})           & $A$ & $B$  & $C$  & \\
& $0.51$ & $0.65$  & $-2.56$ &      \\ \hline
Forced turb. (\citetalias{Lim2024})          & $A'$ & $B'$  & $C'$  & $D'$ \\
$\alpha_{\rm D}$ & $0.15$ & $-0.24$ & $-1.48$ & $1.18$ \\
$\alpha_z$ & $0.13$ & $-0.28$ & $-1.74$ & $1.09$ \\ \hline\hline
\end{tabular}
\end{table}

A comparison between the fits from \citetalias{Lim2024} and \citetalias{Lim2026} shows that forced turbulence significantly weakens SI-driven clumping, yielding supersolar $Z_{\rm crit}$ even in the case of relatively weak turbulence ($\alpha_{\rm D}=10^{-4}$). However, the turbulence used in the simulations of \citetalias{Lim2024} is isotropic and uncorrelated in time, unlike turbulence driven by several
(magneto)hydrodynamical processes that can operate in protoplanetary disks, such as the vertical shear instability (VSI; \citealt{Nelson2013}) and the magnetorotational instability (MRI; \citealt{Velikhov1959,Chandrasekhar1960,BalbusHawley1991}). In this work, we focus on MRI-driven turbulence, while referring to treatments of SI in VSI by \citet{SchaferJohansen2022} and \citet{Schafer2025}.

In the regime of ideal magnetohydrodynamics (MHD), where the gas is sufficiently ionized and well coupled to the magnetic field, the MRI drives strong turbulence that efficiently redistributes angular momentum and facilitates disk accretion. For the most part, however, protoplanetary disks are expected to be weakly ionized, leading to three nonideal MHD effects -- ohmic resistivity, the Hall effect, and ambipolar diffusion (AD) -- which dominate different regions of the disk. Ohmic resistivity and AD are both dissipative and act to stabilize the MRI. In the midplane of the disk where the SI occurs,  Ohmic resistivity dampens turbulence in the inner disk, and AD dampens turbulence in the outer disk \citep{BaiStone2011}. Strong clumping has been observed in simulations with ideal MHD \citep{Johansen2007,Johansen2011}, Ohmic resistivity \citep{Yang2018}, and AD \citep{XuBai2022,HuangBai2025}. Our study focuses on planetesimal formation in the outer disk, where AD is the dominant nonideal MHD effect and MRI is the primary driver of turbulence. (The SI itself also drives turbulence, but only weakly.)

\citet{XuBai2022} found that the inclusion of modest turbulence driven by the MRI with AD leads to lower $Z_{\rm crit}$ for clumping compared to pure SI simulations. This is due to the spontaneous formation of long-lived density and pressure variations by the MRI turbulence -- so-called zonal flows \citep{Johansen2009a} -- in which particles concentrate. They found that concentration inside zonal flows and dust back-reaction work in tandem to generate strong particle clumping. The study by \citet{XuBai2022} was limited to a single $\mathrm{St}$ and a single magnetic field strength. In our study, we extend their work by performing simulations with multiple values of $\mathrm{St}$ in the range $0.01$-–$0.1$, and two different magnetic field strengths producing MRI turbulence with  Shakura–Sunyaev viscosity parameters $\alpha_{\rm SS} =$ $10^{-4}$ and $10^{-3}$. For each combination of $\mathrm{St}$ and magnetic field strength, we perform simulations with varying dust-to-gas surface density ratios $Z$, in order to determine the clumping boundary. This constitutes the first parameter study of the SI interacting with nonideal MHD turbulence.

The paper is organized as follows. In Section \ref{sec:method}, we briefly outline the methodology. Section \ref{sec:results} presents our results for $Z_{\rm crit}$, comparisons with the clumping boundaries from \citetalias{Lim2026} and \citetalias{Lim2024}, and the role of zonal flows. In Section \ref{sec:discussion}, we discuss other types of turbulence and the limitations of the model. Finally, Section \ref{sec:conclusion} summarizes the main conclusions. 

\section{Method}\label{sec:method}
We use the local shearing-box approximation \citep{GoldreichLyndenBell1965} to simulate the evolution of gas and particles in a small portion of a protoplanetary disk. Our 3D simulations include ambipolar diffusion and particle backreaction on the gas. Our methodology largely follows that of \citet{XuBai2022}, with only minor modifications to the simulation setup. In the following, we provide a brief description of our model; for further details, we refer the reader to \citet{XuBai2022}.

\subsection{Numerical method}

We consider an isothermal, non-self-gravitating, protoplanetary disk. We use the Athena MHD code \citep{Stone2008} to model a small portion of the disk located at an arbitrary orbital radius and corotating with the Keplerian angular frequency $\Omega_{\rm K}$. We ignore vertical gravity in the gas, i.e., the gas is unstratified, and thus the initial gas density $\rho_0$ is uniform. We set the following code units: $c_{\rm s}=\Omega_{\rm K}=\rho_0=1$, where $c_{\rm s}$ is the isothermal sound speed. Given our choice of $c_{\rm s}$ and $\Omega_{\rm K}$, the gas scale height would be $H_{\rm g}=1$. We apply shearing-periodic boundary conditions in the radial ($x$) direction, and periodic boundary conditions in the azimuthal ($y$) and vertical ($z$) directions.

We impose a net vertical magnetic field $B_0$, the strength of which is characterized by the plasma $\beta$, 
\begin{equation}
    \beta_0\equiv \frac{\rho_0c_{\rm s}^2}{B_0^2/2\mu_0},
\end{equation}
which corresponds to the ratio of gas pressure to the magnetic pressure of the net vertical field. In the above equation, $\mu_0$ is the permeability of vacuum, and we adopt the code units $\mu_0 = 1$. The strength of nonideal MHD effects is quantified by the Elsässer number, which for AD is given by
\begin{equation}
    \mathrm{Am}=\frac{\gamma \rho_{\rm i}}{\Omega_{\rm K}},
\end{equation}
where $\gamma$ is the coefficient of momentum exchange in collisions between ions and neutrals, and $\rho_{\rm i}$ is the ion density. The limit $\mathrm{Am}\rightarrow\infty$ corresponds to ideal MHD, where the magnetic field is perfectly coupled to the gas. 

Dust is modeled as Lagrangian particles and coupled to the gas via aerodynamic drag \citep{BaiStone2010_oct}. The degree of aerodynamic coupling between gas and particles is characterized by St. We note that dust can also be modeled as a pressureless fluid; however, fluid-based models have been shown to reach lower peak dust densities for the SI than particle-based simulations \citep{Baronett2026}. The difference decreases with increasing resolution but is not completely eliminated even at $1024^2$ resolution in axisymmetric, unstratified SI simulations. Since the computational cost of 3D SI simulations with non-ideal MHD at such high resolution is prohibitive for a parameter study, we adopt particle-based models that can be run at lower resolutions.
Unlike particles, gas in protoplanetary disks is pressure-supported and orbits at a sub-Keplerian velocity, causing the particles to experience a headwind and drift radially toward the star. Rather than implementing the effects of the radial pressure gradient directly on the gas, a force is added to the particle equations of motion to mimic the outwardly decreasing radial pressure gradient of the gas. The pressure gradient is characterized by $\Pi \equiv \eta v_{\rm K}/c_{\rm s}$, where $\eta v_{\rm K}$ is the deviation from the Keplerian velocity $v_{\rm K}$. We include vertical gravity for the particles, allowing them to settle toward the midplane. 

\subsection{Simulation setup}
Our shearing box is mapped onto a fixed, regular grid with a standard size of $(L_x, L_y,L_z)=(0.8,0.8,0.8)H_{\rm g}$. The standard resolution is 250 cells per gas scale height, and thus the number of grid cells in each direction is $(N_x, N_y,N_z)=(200,200,200)$. We also perform higher-resolution models for convergence tests, as listed in Table~\ref{table: simulation list}. We adopt the same radial pressure gradient as in \citetalias{Lim2024} and \citetalias{Lim2026}, $\Pi=0.05$. Throughout this study, the Elsässer number is set to $\mathrm{Am}=2$, following \citet{XuBai2022}, and two magnetic field strengths are considered, $\beta_0=6\times10^{4}$ and $\beta_0=6\times10^{3}$. The corresponding wavelengths of fastest MRI growth are $0.039H_{\rm g}$ and $0.122H_{\rm g}$, which gives a minimum resolution of $10$ cells per wavelength. Particles in our simulations have fixed Stokes numbers. We explore $\mathrm{St}=0.01$, $0.03$, and $0.1$. 

Our aim is to identify the conditions under which the SI leads to strong clumping in a disk with self-consistently driven MRI turbulence. We adopt the same strong clumping criterion as \citet{LiYoudin2021}, \citetalias{Lim2024}, and \citetalias{Lim2026}: 
\begin{equation}\label{eq: Hill density}
    \rho_{p,\rm{max}}>\rho_{\rm H}=9\rho_0\sqrt{\frac{\pi}{8}}Q\simeq 180\rho_0,
\end{equation}
where $\rho_{p,\rm{max}}$ is the maximum particle density, $\rho_{\rm H}$ is the Hill density, and $Q$ is the Toomre parameter \citep{Toomre1964}. The Toomre parameter is set to $Q=32$ (implicitly fixing the disk mass or the box location, though we do not actually compute gravity), consistent with previous work; this choice corresponds to a low-mass disk and yields a conservative threshold for strong clumping.

We perform simulations with varying dust-to-gas surface density ratios $Z$ in order to determine $Z_{\rm crit}$ for each combination of $\beta_0$ and $\mathrm{St}$. Each simulation is initialized without particles and evolved for $\sim 30$ orbital periods to allow the MRI to develop and fully saturate into turbulence. Particles are then inserted with random radial and azimuthal positions and a Gaussian vertical distribution with a particle scale height $H_{\rm p}$, which is varied between simulations. The initial gas and particle velocities follow the \citet{Nakagawas1986} equilibrium solution. In our standard models, we use $200^3$ particles per simulation, corresponding to an average of one particle per grid cell, a resolution we maintain in our higher resolution models. After insertion, particles settle toward the midplane. Simulations are evolved for just under 500 orbital periods, or until the Hill density is clearly exceeded, except in a few cases where they are extended to probe the late-time evolution. The full set of simulations is shown in Table \ref{table: simulation list}.

\begin{table*}[]
\centering
\begin{tabular}{llllllll}
\hline\hline
St & $Z$ & $\beta_0$ & $L_x\times L_y\times L_z / H^3$ & $N_x\times N_y\times N_z$ & $t_{\rm sim}/P$ & $\max (\rho_p)/\rho_0$ & Strong clumping? \\
\hline

$1\mathrm{e}{-2}$ & $1.0\mathrm{e}{-2}$  & $6\mathrm{e}{4}$ & $0.8\times0.8\times0.8$ & $200\times200\times200$ & $637$ & $6.3\mathrm{e}{0}$   & N \\
$1\mathrm{e}{-2}$ & $2.0\mathrm{e}{-2}$  & $6\mathrm{e}{4}$ & $0.8\times0.8\times0.8$ & $200\times200\times200$ & $637$ & $8.22\mathrm{e}{1}$  & N \\
$1\mathrm{e}{-2}$ & $3.5\mathrm{e}{-2}$ & $6\mathrm{e}{4}$ & $0.8\times0.8\times0.8$ & $200\times200\times200$ & $477$ & $5.34\mathrm{e}{2}$ & Y \\
$1\mathrm{e}{-2}$ & $3.5\mathrm{e}{-2}$ & $6\mathrm{e}{4}$ & $1.4\times0.8\times0.8$ & $350\times200\times200$ & $300$ & $2.54\mathrm{e}{2}$ & Y \\
$1\mathrm{e}{-2}$ & $3.5\mathrm{e}{-2}$ & $6\mathrm{e}{4}$ & $0.8\times0.8\times0.8$ & $400\times400\times400$ & $450$ & $2.25\mathrm{e}{2}$ & Y \\
$1\mathrm{e}{-2}$ & $5.0\mathrm{e}{-2}$  & $6\mathrm{e}{4}$ & $0.8\times0.8\times0.8$ & $200\times200\times200$ & $555$ & $2.58\mathrm{e}{2}$ & Y \\
$1\mathrm{e}{-2}$ & $5.0\mathrm{e}{-2}$  & $6\mathrm{e}{4}$ & $0.8\times0.8\times0.8$ & $400\times400\times400$ & $142$ & $2.95\mathrm{e}{2}$ & Y \\
\hline

$3\mathrm{e}{-2}$ & $5.0\mathrm{e}{-3}$  & $6\mathrm{e}{4}$ & $0.8\times0.8\times0.8$ & $200\times200\times200$ & $477$ & $1.06\mathrm{e}{1}$  & N \\
$3\mathrm{e}{-2}$ & $7.5\mathrm{e}{-3}$ & $6\mathrm{e}{4}$ & $0.8\times0.8\times0.8$ & $200\times200\times200$ & $557$ & $1.19\mathrm{e}{2}$ & N \\
$3\mathrm{e}{-2}$ & $1.0\mathrm{e}{-2}$   & $6\mathrm{e}{4}$ & $0.8\times0.8\times0.8$ & $200\times200\times200$ & $477$ & $5.02\mathrm{e}{2}$ & Y \\
$3\mathrm{e}{-2}$ & $3.0\mathrm{e}{-2}$   & $6\mathrm{e}{4}$ & $0.8\times0.8\times0.8$ & $200\times200\times200$ & $97$  & $7.94\mathrm{e}{2}$ & Y \\
\hline

$1\mathrm{e}{-1}$ & $1.0\mathrm{e}{-3}$  & $6\mathrm{e}{4}$ & $0.8\times0.8\times0.8$ & $200\times200\times200$ & $637$ & $4.57\mathrm{e}{1}$   & N \\
$1\mathrm{e}{-1}$ & $2.5\mathrm{e}{-3}$ & $6\mathrm{e}{4}$ & $0.8\times0.8\times0.8$ & $200\times200\times200$ & $477$ & $4.39\mathrm{e}{2}$  & Y \\
$1\mathrm{e}{-1}$ & $2.5\mathrm{e}{-3}$ & $6\mathrm{e}{4}$ & $1.4\times0.8\times0.8$ & $350\times200\times200$ & $172$ & $6.10\mathrm{e}{2}$  & Y \\
$1\mathrm{e}{-1}$ & $2.5\mathrm{e}{-3}$ & $6\mathrm{e}{4}$ & $0.8\times0.8\times0.8$ & $400\times400\times400$ & $169$ & $1.76\mathrm{e}{3}$ & Y \\
$1\mathrm{e}{-1}$ & $5.0\mathrm{e}{-3}$  & $6\mathrm{e}{4}$ & $0.8\times0.8\times0.8$ & $200\times200\times200$ & $157$ & $1.98\mathrm{e}{3}$ & Y \\
$1\mathrm{e}{-1}$ & $5.0\mathrm{e}{-3}$  & $6\mathrm{e}{4}$ & $1.4\times0.8\times0.8$ & $350\times200\times200$ & $175$ & $3.04\mathrm{e}{3}$ & Y \\
$1\mathrm{e}{-1}$ & $5.0\mathrm{e}{-3}$  & $6\mathrm{e}{4}$ & $0.8\times0.8\times0.8$ & $400\times400\times400$ & $67$  & $5.05\mathrm{e}{2}$  & Y \\
$1\mathrm{e}{-1}$ & $7.5\mathrm{e}{-3}$ & $6\mathrm{e}{4}$ & $0.8\times0.8\times0.8$ & $200\times200\times200$ & $67$  & $2.79\mathrm{e}{3}$ & Y \\
\hline

$1\mathrm{e}{-2}$ & $5.0\mathrm{e}{-2}$   & $6\mathrm{e}{3}$ & $0.8\times0.8\times0.8$ & $200\times200\times200$ & $637$ & $1.38\mathrm{e}{1}$  & N \\
$1\mathrm{e}{-2}$ & $7.5\mathrm{e}{-2}$  & $6\mathrm{e}{3}$ & $0.8\times0.8\times0.8$ & $200\times200\times200$ & $477$ & $2.44\mathrm{e}{2}$ & Y \\
$1\mathrm{e}{-2}$ & $7.5\mathrm{e}{-2}$  & $6\mathrm{e}{3}$ & $1.4\times0.8\times0.8$ & $350\times200\times200$ & $239$ & $3.27\mathrm{e}{2}$ & Y \\
$1\mathrm{e}{-2}$ & $7.5\mathrm{e}{-2}$  & $6\mathrm{e}{3}$ & $0.8\times0.8\times0.8$ & $400\times400\times400$ & $407$ & $3.1\mathrm{e}{1}$ & N \\
$1\mathrm{e}{-2}$ & $1.0\mathrm{e}{-1}$    & $6\mathrm{e}{3}$ & $0.8\times0.8\times0.8$ & $200\times200\times200$ & $477$ & $8.08\mathrm{e}{2}$ & Y \\
$1\mathrm{e}{-2}$ & $1.0\mathrm{e}{-1}$    & $6\mathrm{e}{3}$ & $0.8\times0.8\times0.8$ & $400\times400\times400$ & $173$ & $3.69\mathrm{e}{2}$ & Y \\
$1\mathrm{e}{-2}$ & $1.5\mathrm{e}{-1}$   & $6\mathrm{e}{3}$ & $0.8\times0.8\times0.8$ & $200\times200\times200$ & $131$ & $5.82\mathrm{e}{2}$ & Y \\
\hline

$3\mathrm{e}{-2}$ & $2.0\mathrm{e}{-2}$ & $6\mathrm{e}{3}$ & $0.8\times0.8\times0.8$ & $200\times200\times200$ & $477$ & $2.4\mathrm{e}{0}$   & N \\
$3\mathrm{e}{-2}$ & $4.0\mathrm{e}{-2}$ & $6\mathrm{e}{3}$ & $0.8\times0.8\times0.8$ & $200\times200\times200$ & $477$ & $5.08\mathrm{e}{1}$  & N \\
$3\mathrm{e}{-2}$ & $5.0\mathrm{e}{-2}$ & $6\mathrm{e}{3}$ & $0.8\times0.8\times0.8$ & $200\times200\times200$ & $477$ & $6.08\mathrm{e}{2}$ & Y \\
$3\mathrm{e}{-2}$ & $6.0\mathrm{e}{-2}$ & $6\mathrm{e}{3}$ & $0.8\times0.8\times0.8$ & $200\times200\times200$ & $477$ & $5.77\mathrm{e}{2}$ & Y \\
\hline

$1\mathrm{e}{-1}$ & $1.0\mathrm{e}{-2}$  & $6\mathrm{e}{3}$ & $0.8\times0.8\times0.8$ & $200\times200\times200$ & $477$  & $7.26\mathrm{e}{1}$   & N \\
$1\mathrm{e}{-1}$ & $1.5\mathrm{e}{-2}$ & $6\mathrm{e}{3}$ & $0.8\times0.8\times0.8$ & $200\times200\times200$ & $477$  & $1.51\mathrm{e}{3}$ & Y \\
$1\mathrm{e}{-1}$ & $1.5\mathrm{e}{-2}$ & $6\mathrm{e}{3}$ & $1.4\times0.8\times0.8$ & $350\times200\times200$ & $174$  & $3.04\mathrm{e}{3}$ & Y \\
$1\mathrm{e}{-1}$ & $1.5\mathrm{e}{-2}$ & $6\mathrm{e}{3}$ & $0.8\times0.8\times0.8$ & $400\times400\times400$ & $119$  & $8.90\mathrm{e}{2}$  & Y \\
$1\mathrm{e}{-1}$ & $2.0\mathrm{e}{-2}$  & $6\mathrm{e}{3}$ & $0.8\times0.8\times0.8$ & $200\times200\times200$ & $477$  & $3.57\mathrm{e}{3}$ & Y \\

\hline\hline
\end{tabular}
\caption{List of simulations and maximum particle density ($\max (\rho_p)/\rho_0$)) obtained during the simulation time ($t_{\rm sim}$). Simulations in which $\max (\rho_p)/\rho_0>180$ are classified as strong clumping.}
\label{table: simulation list}
\end{table*}

\section{Results}\label{sec:results}
\begin{figure*}
    \centering
    \includegraphics[width=.98\textwidth]{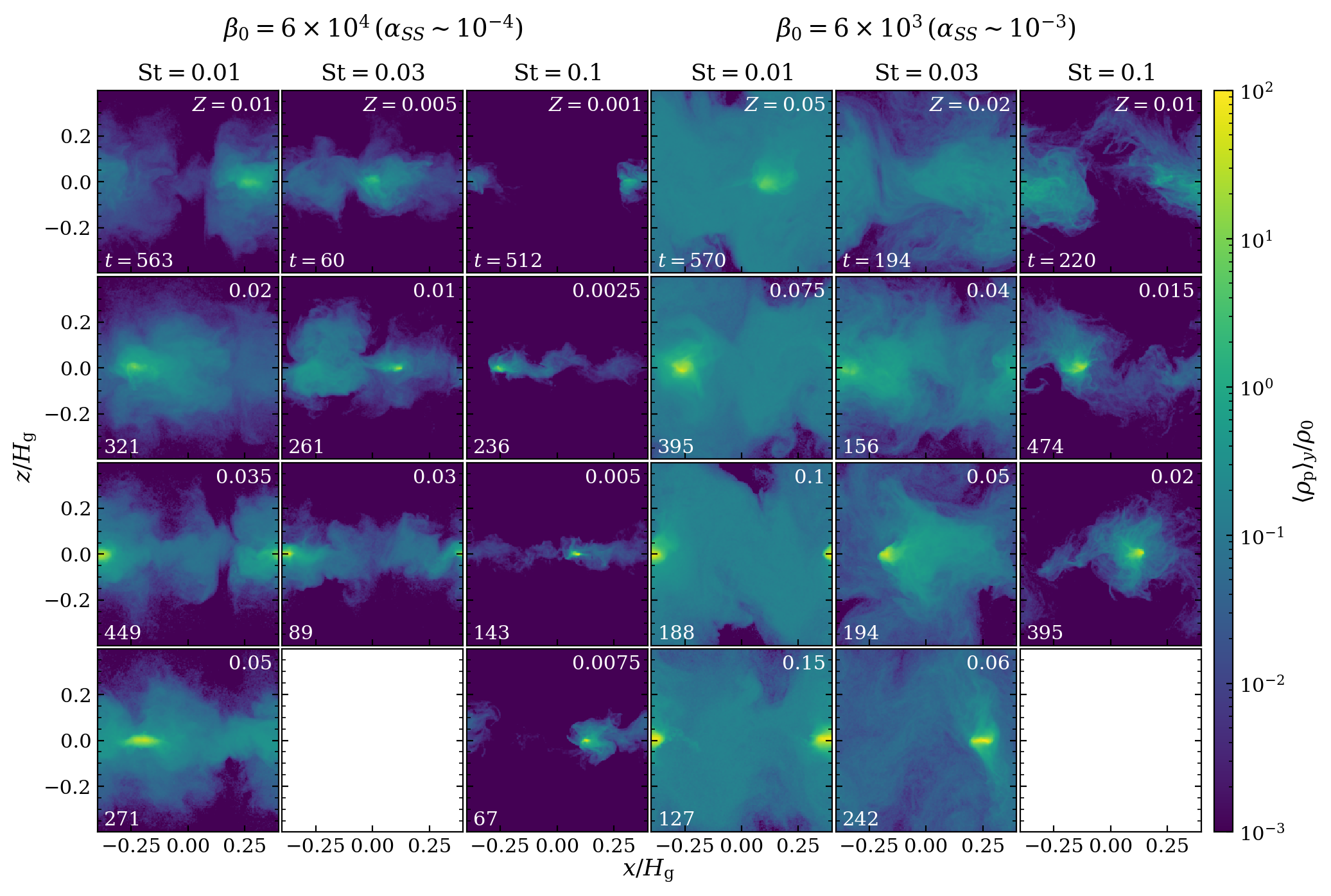}
    \caption{Images of the azimuthally averaged particle density for all simulations at fiducial resolution ($250/H_{\rm g}$) and box-size ($0.8H_{\rm g}\times0.8H_{\rm g}\times0.8H_{\rm g}$) in the parameter study. For each simulation, we show the time when the maximum dust-to-gas ratio occurs (indicated in the bottom-left corner of each panel). The three columns on the left correspond to simulations with $\beta_0=6\times10^4$, yielding $\alpha_{\rm SS}\sim10^{-4}$, while the three columns on the right correspond to simulations with $\beta_0=6\times10^3$, yielding $\alpha_{\rm SS}\sim10^{-3}$. The dust-to-gas surface density ratio $Z$ of each simulation is indicated in the top-right corner of each panel.}
    \label{fig:zVSx_rhop}
\end{figure*}

Figure \ref{fig:zVSx_rhop} shows images of the azimuthally averaged particle density for all simulations in the parameter study. For each simulation, the time chosen is when the maximum dust-to-gas ratio occurs. The high-density clumps arise from a combination of particle trapping in zonal flows and concentration due to dust back-reaction (see Sect.~\ref{sec: results zonal flows}). Runs with higher dust-to-gas surface density $Z$ typically produce denser clumps. We note that the clumps seen in the $x$-$z$ plane are in fact high-density filaments that often extend across the entire $y$-extent of the simulation domain. The impact of MRI turbulence is evident in the dust layer, which exhibits a highly irregular structure. Simulations with smaller $\mathrm{St}$ have larger dust scale heights and exhibit larger, more diffuse clumps.

Simulations with lower plasma $\beta$ exhibit stronger turbulence, leading to larger dust scale heights and less smooth dust layers. The strength of turbulence can be characterized in several ways, with one commonly used measure being the Shakura–Sunyaev stress parameter ($\alpha_{\rm SS}$; \citealt{ShakuraSunyaev1973}). The parameter $\alpha_{\rm SS}$ quantifies the rate of angular momentum transport in the disk and can be evaluated as
\begin{equation}\label{eq: alpha_ss}
    \alpha_{\rm SS} \equiv \frac{-\langle B_xB_y\rangle/\mu_0+\langle\rho_0 u_xu_y\rangle}{\rho_0 c_{\rm s}^2},
\end{equation}
where $B$ is the magnetic field strength and $u$ is the gas velocity (e.g. \citealt{Brandenburg1998}). The first term represents the Maxwell stress, while the second corresponds to the Reynolds stress; the total stress is normalized by the thermal pressure. We measure $\alpha_{\rm SS}$ in simulations without dust feedback and obtain approximately $\alpha_{\rm SS}\simeq10^{-4}$ and $\alpha_{\rm SS}\simeq10^{-3}$ for $\beta_0=6\times10^4$ and $\beta_0=6\times10^3$, respectively (see Table \ref{table: alpha}).  

\subsection{Threshold for strong clumping}
\begin{figure*}
    \centering
    \includegraphics[width=.98\textwidth]{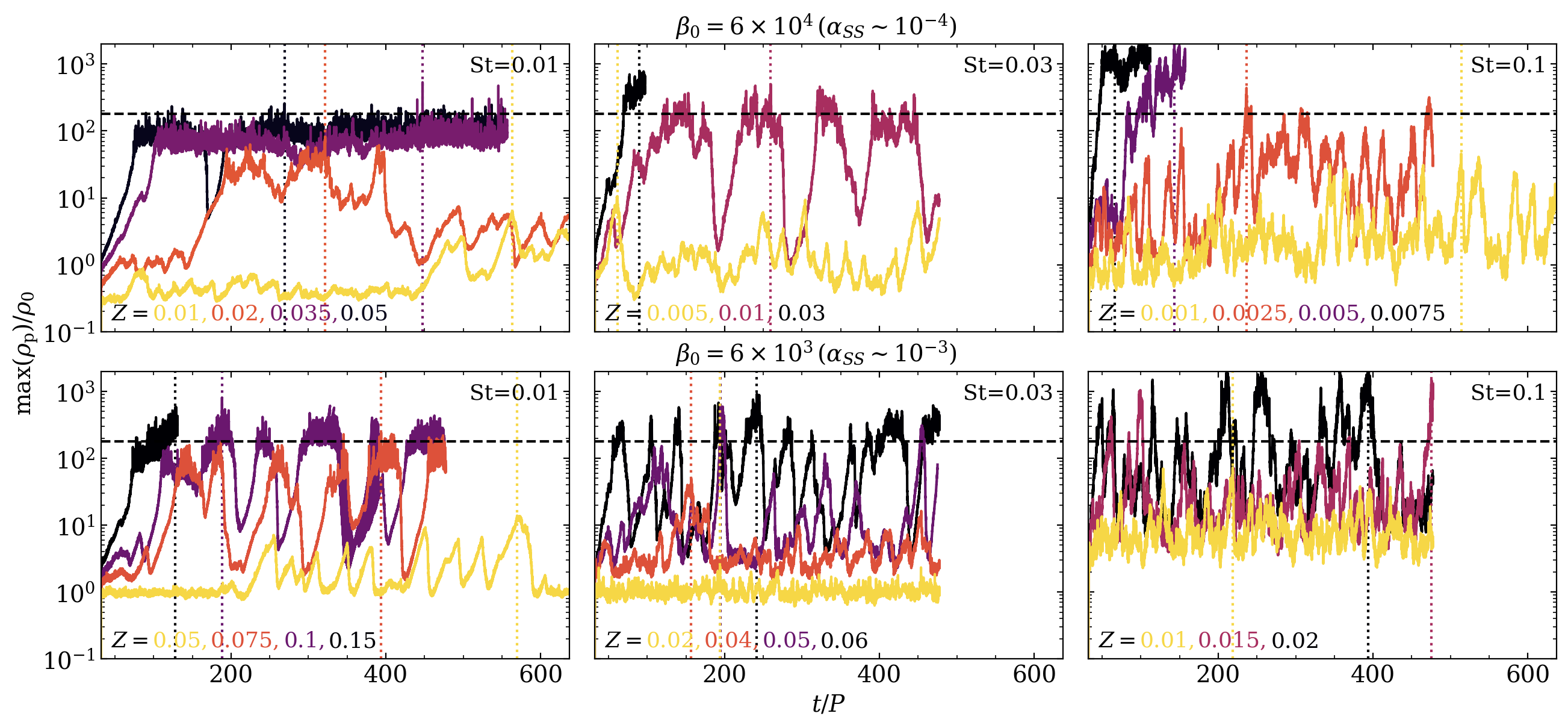}
\caption{Maximum particle density as a function of time in terms of orbital period $P$ for the same simulations as in Figure \ref{fig:zVSx_rhop}. The dashed line indicates the Hill density (Eq.~\eqref{eq: Hill density}), the dotted lines mark the time when the maximum dust-to-gas ratio occurs, and the dust-to-gas surface density ratio $Z$ for each simulation is shown in the bottom-left corner of each panel. }
    \label{fig:d2g}
\end{figure*}

\begin{figure*}
    \centering
    \includegraphics[width=\textwidth]{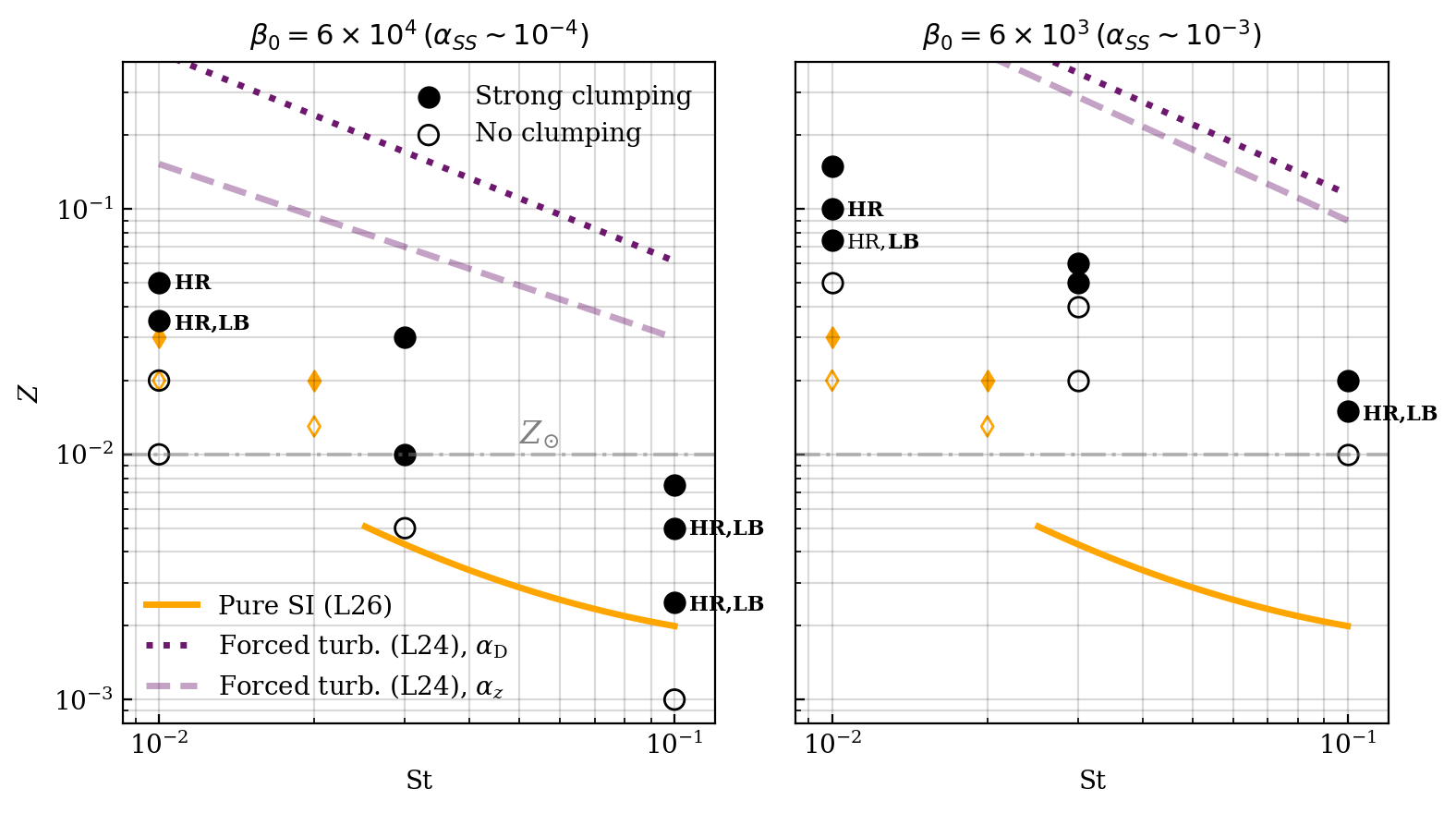}
    \caption{Overview of all simulations in the parameter study, showing whether they exhibit strong clumping (filled circles) or not (empty circles). Runs that have been rerun at double the spatial resolution and with a 75\% longer radial box-length are marked HR and LB, respectively, while runs that exhibit strong clumping are marked with boldface. Results from the 3D pure SI simulations by \citetalias{Lim2026} are shown as a solid orange line and diamond markers, where filled markers indicate strong clumping. The dashed and dotted purple lines show the comparison with \citetalias{Lim2024}, who performed 3D SI simulations including isotropicaly forced turbulence. Our simulations with relatively weak turbulence (left panel) show $Z_{\rm crit}$ similar to the non-turbulent cases, whereas stronger turbulence (right panel) results in higher $Z_{\rm crit}$, although still significantly smaller than in the isotropically forced-turbulence models.}
    \label{fig:Zcrit}
\end{figure*}

Figure \ref{fig:d2g} shows the time evolution of the maximum particle density for all simulations in the parameter study. Simulations in which the maximum density exceeds the Hill density (dashed line; Eq. \ref{eq: Hill density}) are classified as exhibiting strong clumping and are shown in Figure \ref{fig:Zcrit} as solid circles. We emphasize that strong clumping is not necessarily synonymous with SI clumping, which is typically identified by the presence of distinct clumps in the $z$–$x$ plane and a sharp increase in maximum particle density. For example, the simulation with $\beta_0=6\times10^4$, $\mathrm{St}=0.01$, and $Z=0.02$ shows clear signatures of SI clumping, but does not satisfy our criterion for strong clumping.

Since we do not model particle self-gravity, no actual planetesimals form in the simulations, and the dense clumps and filaments that develop are often transient. As in previous parameter studies, $Z_{\rm crit}$ decreases with increasing $\mathrm{St}$, at least over the range of $\mathrm{St}$ considered here. In simulations with relatively weak turbulence $\alpha_{\rm SS}\sim10^{-4}$, strong clumping is observed for dust-to-gas ratios as low as $Z=0.25Z_{\odot}$, where $Z_{\odot}=0.01$ is the solar dust-to-gas ratio. For $\mathrm{St}<0.03$, supersolar dust-to-gas ratios are required. In simulations with stronger turbulence $\alpha_{\rm SS}\sim10^{-3}$, supersolar dust-to-gas ratios are required regardless of $\mathrm{St}$. In this regime, particle concentration by dust back-reaction and MRI zonal flows is insufficient for planetesimal formation to occur with a Solar dust-to-gas ratio and particles with a single, fixed Stokes number. A proper treatment of dust growth could change this, as recent studies have shown that dust clumping by the SI and dust growth boost each other \citep{Ho2024,Carrera2025,Tominaga2025,Vallucci-Goy2026,Pierens2026}.

Some previous parameter studies, including \citetalias{Lim2024} and \citetalias{Lim2026}, also present the clumping boundary in terms of the midplane dust-to-gas volume density ratio $\epsilon$. Unlike $Z$, which is a simulation input parameter, $\epsilon$ must be measured from the simulations. To obtain a reliable value of $\epsilon$, it needs to be measured during the pre-clumping phase, defined as the interval after an equilibrium dust layer has formed but before significant clumps or filaments appear; this is also the phase during which the particle scale height should be measured. In our study, a significant fraction of the simulations do not exhibit a well-defined pre-clumping phase. As a result, we are unable to apply the same procedure to obtain a reliable estimate of $\epsilon$ or $H_{\rm p}$. Therefore, we choose not to present the clumping boundary in terms of $\epsilon$.

\subsection{Comparison with pure SI and isotropically forced-turbulence studies}\label{subsec: results compare LI24/LI26}
The solid orange line and diamond markers in Fig. \ref{fig:Zcrit} show the results of the pure SI simulations in 3D from \citetalias{Lim2026} (Eq. \ref{eq: L25}). Our $Z_{\rm crit}$ agrees well with the results of \citetalias{Lim2026} in the case of relatively weak MRI turbulence. This suggests that MRI-driven turbulence with AD does not hinder planetesimal formation, provided that the turbulence is not too strong. When the turbulence strength is increased by a factor of ten (right panel of Fig. \ref{fig:Zcrit}), $Z_{\rm crit}$ becomes several times higher than in the weak-turbulence scenario. Strong MRI turbulence thus weakens particle concentration.

We note that our comparison with \citetalias{Lim2026} is not a one-to-one comparison, since the simulation setups are quite different. The lack of external turbulence in \citetalias{Lim2026} allows them to adopt a smaller vertical box height, and they use a fiducial domain size of $(0.8\times0.2\times0.2)H_{\rm g}$, whereas we adopt a fiducial domain size of ($0.8\times0.8\times0.8)H_{\rm g}$. They also use a higher fiducial resolution of $640H_{\rm g}^{-1}$, compared to our fiducial resolution of $250H_{\rm g}^{-1}$. Importantly, a convergence study with simulations at $500H_{\rm g}^{-1}$ shows that our main results remain unchanged at higher resolution (see Sect.~\ref{subsec:convergence}). The lower resolution in our study is a compromise to the increased computational cost associated with the larger domain size and the inclusion of MHD. 

Finally, \citetalias{Lim2026} adopts outflow boundary conditions in the vertical direction.
In SI simulations, particle–gas interactions near the midplane can drive gas perturbations away from the midplane, which may reflect and induce additional particle stirring when periodic or reflecting boundary conditions are used \citep{Li2018}. This effect becomes more alleviated as the vertical domain-size is increased, as shown by \citet{Lim2025} who compared runs with $L_z=0.2\, H_{\rm g}$ and $L_z=0.4\, H_{\rm g}$ with outflow and periodic boundary conditions. In our work, similarly to \citetalias{Lim2024}, we assume that turbulence generated by the MRI dominates over any additional stirring introduced by the vertical boundary conditions.

\begin{figure}
    \centering
    \includegraphics[width=\columnwidth]{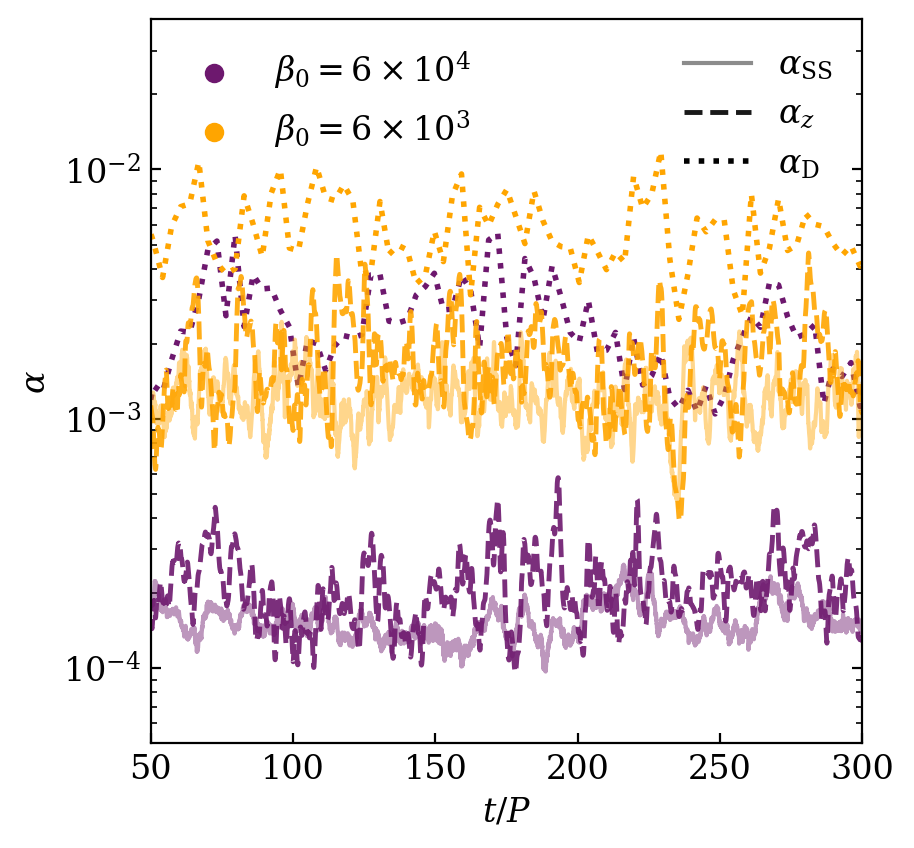}
\caption{Time evolution of the three turbulent parameters $\alpha_\mathrm{SS}$ (Eq.~\eqref{eq: alpha_ss}), $\alpha_z$ (Eq.~\eqref{eq: alpha_Z}), and $\alpha_\mathrm{D}$ (Eq.~\eqref{eq: alpha_D}), measured in simulations without particle feedback and with $\mathrm{St}=0.1$.}
    \label{fig:alpha}
\end{figure}

\begin{table*}[]
\centering
\begin{tabular}{lllllllllll}
\hline\hline
$\beta_0$ & Sim & $H_{\rm p}$ & $\alpha_z$ & $\delta u_x^{'}$ & $\delta u_y^{'}$ & $\delta u_z^{'}$ & $\alpha_{\rm D}$ & $-B_xB_y$ & $\rho_0 u_xu_y$ & $\alpha_{\rm SS}$ \\
\hline
$6\mathrm{e}{4}$ & Fid & $4.576\mathrm{e}{-2}$ & $2.150\mathrm{e}{-4}$ & $1.382\mathrm{e}{-2}$ & $3.552\mathrm{e}{-2}$ & $1.232\mathrm{e}{-2}$ & $2.390\mathrm{e}{-3}$ & $1.218\mathrm{e}{-4}$ & $3.535\mathrm{e}{-5}$ & $1.571\mathrm{e}{-4}$ \\
$6\mathrm{e}{4}$ & HR & $4.589\mathrm{e}{-2}$ & $1.996\mathrm{e}{-4}$ & $1.380\mathrm{e}{-2}$ & $2.198\mathrm{e}{-2}$ & $1.248\mathrm{e}{-2}$ & $1.084\mathrm{e}{-3}$ & $1.370\mathrm{e}{-4}$ & $3.612\mathrm{e}{-5}$ & $1.764\mathrm{e}{-4}$ \\
$6\mathrm{e}{4}$ & LB & $4.427\mathrm{e}{-2}$ & $2.192\mathrm{e}{-4}$ & $1.379\mathrm{e}{-2}$ & $3.613\mathrm{e}{-2}$ & $1.227\mathrm{e}{-2}$ & $2.768\mathrm{e}{-3}$ & $1.262\mathrm{e}{-4}$ & $3.620\mathrm{e}{-5}$ & $1.624\mathrm{e}{-4}$ \\
\hline
$6\mathrm{e}{3}$ & Fid & $1.275\mathrm{e}{-1}$ & $1.723\mathrm{e}{-3}$ & $3.786\mathrm{e}{-2}$ & $4.053\mathrm{e}{-2}$ & $3.508\mathrm{e}{-2}$ & $5.920\mathrm{e}{-3}$ & $9.216\mathrm{e}{-4}$ & $3.147\mathrm{e}{-4}$ & $1.236\mathrm{e}{-3}$ \\
$6\mathrm{e}{3}$ & HR & $1.380\mathrm{e}{-1}$ & $2.021\mathrm{e}{-3}$ & $3.736\mathrm{e}{-2}$ & $3.884\mathrm{e}{-2}$ & $3.615\mathrm{e}{-2}$ & $6.824\mathrm{e}{-3}$ & $9.598\mathrm{e}{-4}$ & $3.085\mathrm{e}{-4}$ & $1.268\mathrm{e}{-3}$ \\
$6\mathrm{e}{3}$ & LB & $1.646\mathrm{e}{-1}$ & $2.859\mathrm{e}{-3}$ & $3.731\mathrm{e}{-2}$ & $4.245\mathrm{e}{-2}$ & $4.009\mathrm{e}{-2}$ & $8.444\mathrm{e}{-3}$ & $9.236\mathrm{e}{-4}$ & $3.049\mathrm{e}{-4}$ & $1.229\mathrm{e}{-3}$ \\
\hline\hline
\end{tabular}
\caption{Time-averaged quantities measured from simulations performed without particle feedback and using $\mathrm{St}=0.1$. Three different grids were used: "Fid" corresponds to the fiducial domain size and resolution; "HR" corresponds to runs performed with two times higher spatial resolution ($500/H_{\rm g}$); and "LB" corresponds to runs performed with a 75\% longer radial domain size ($L_x=1.4 H_{\rm g}$).}
\label{table: alpha}
\end{table*}

To compare with the SI simulations of \citetalias{Lim2024} that include isotropically forced turbulence, we first need to calculate the turbulence parameters $\alpha_{\rm D}$ and $\alpha_{Z}$ used in that study. To this end, we perform simulations without particle feedback and with $\mathrm{St}=0.1$. The parameter $\alpha_{Z}$ represents gaseous diffusion in the vertical direction and is obtained by assuming a balance between vertical settling and turbulent diffusion of the dust particles driven by the gas turbulence:
\begin{equation} \label{eq: alpha_Z}
\alpha_z = \textrm{St}\left[\frac{(H_{\rm p}/H_{\rm g})^2}{1-(H_{\rm p}/H_{\rm g})^2}\right].
\end{equation}
The particle scale heights are obtained by calculating the standard deviation of the particles' vertical position, after confirming that $\rho_{\rm p}(z)$ exhibits an approximately Gaussian profile. The values reported in Table~\ref{table: alpha} have been averaged over 300 orbital periods. The time evolution of $\alpha_{z}$ is shown as dashed lines in Fig. \ref{fig:alpha}, and the time averages of $H_{\rm p}$ and $\alpha_{z}$ are provided in Table \ref{table: alpha}.

The parameter $\alpha_{\rm D}$, which represents the total spatial diffusion of turbulent gas, can be related to $\alpha_z$ via
\begin{equation} \label{eq: alpha_D}
\alpha_{\rm D} = \alpha_z (l^2 + m^2 + 1),
\end{equation}
where $l\equiv \delta u’_x/\delta u’_z$ and $m\equiv \delta u’_y/\delta u’_z$ (see derivation in \citetalias{Lim2024}, eq. 13-15). The velocity fluctuations of the $i$th component are calculated as
\begin{equation}
\delta u’_i \equiv \sqrt{\langle u_i^{\prime 2} \rangle - \langle u_i^{\prime} \rangle^2},
\end{equation}
where $u_i^{\prime}$ represents the gas velocity after subtraction of the Keplerian shear. 
The time evolution of $\alpha_{\rm D}$ is shown in Fig. \ref{fig:alpha} as dotted lines, and the time-averaged values of $\alpha_{\rm D}$ and the components of $\delta u’$ are provided in Table \ref{table: alpha}.

In the case of isotropic turbulence, $\alpha_{\rm D}=3\alpha_z$. The MRI turbulence in our strong-field simulation with $\beta_0=6\times 10^3$ is close to isotropic, with $\alpha_{\rm D}\approx 3.5\alpha_z$. In contrast, the MRI turbulence in the weak-field simulation with $\beta_0=6\times 10^4$ is highly anisotropic, with $\alpha_{\rm D}\gg 3\alpha_z$. A comparison of the components of the velocity fluctuations reveals that the azimuthal velocity fluctuations are significantly stronger than those in the radial and vertical directions. 

We use the time-averaged values of $\alpha_{z}$ and $\alpha_{\rm D}$ in the fit of \citetalias{Lim2024} (Eq.~\ref{eq: L24}) and obtain the dashed and dotted lines in Fig.~\ref{fig:Zcrit}. Regardless of which of the two turbulence measures is used in the comparison, we obtain significantly lower $Z_{\rm crit}$ for strong clumping in our study, indicating that self-consistently generated MRI turbulence poses a smaller obstacle to planetesimal formation than forced turbulence. As discussed in more detail in Section \ref{sec: results zonal flows}, the MRI generates zonal flows, which aid in particle concentration, whereas forced turbulence acts only to diffuse particles. In the case of relatively weak turbulence, the concentrating effect counteracts diffusion, leading to values of $Z_{\rm crit}$ comparable to those obtained in pure SI simulations. As the turbulence strength is increased, however, diffusion becomes dominant and $Z_{\rm crit}$ increases. This is in agreement with the findings of \citetalias{Lim2024}, which find that SI clumping is prevented when turbulence only acts to diffuse dust.

Similar to our comparison with \citetalias{Lim2026}, our comparison with \citetalias{Lim2024} is not a one-to-one comparison. They use a domain of size $L_x=0.4H_{\rm g}$, $L_y=0.2H_{\rm g}$, and $L_z$ between $0.8-2H_{\rm g}$, with larger $L_z$ used in runs with high turbulence and/or small Stokes numbers. They also use a higher resolution of $640H_{\rm g}^{-1}$. Finally, the simulations of \citetalias{Lim2026} include particle self-gravity, and their clumping criterion is based on gravitational collapse. This is a stricter clumping criterion than exceeding the Hill density, because after reaching the Hill density the dust clump also needs to overcome diffusion at smaller scales in order for collapse to occur. 

\subsection{The role of zonal flows}\label{sec: results zonal flows}
\begin{figure*}
    \centering
    \includegraphics[width=\textwidth]{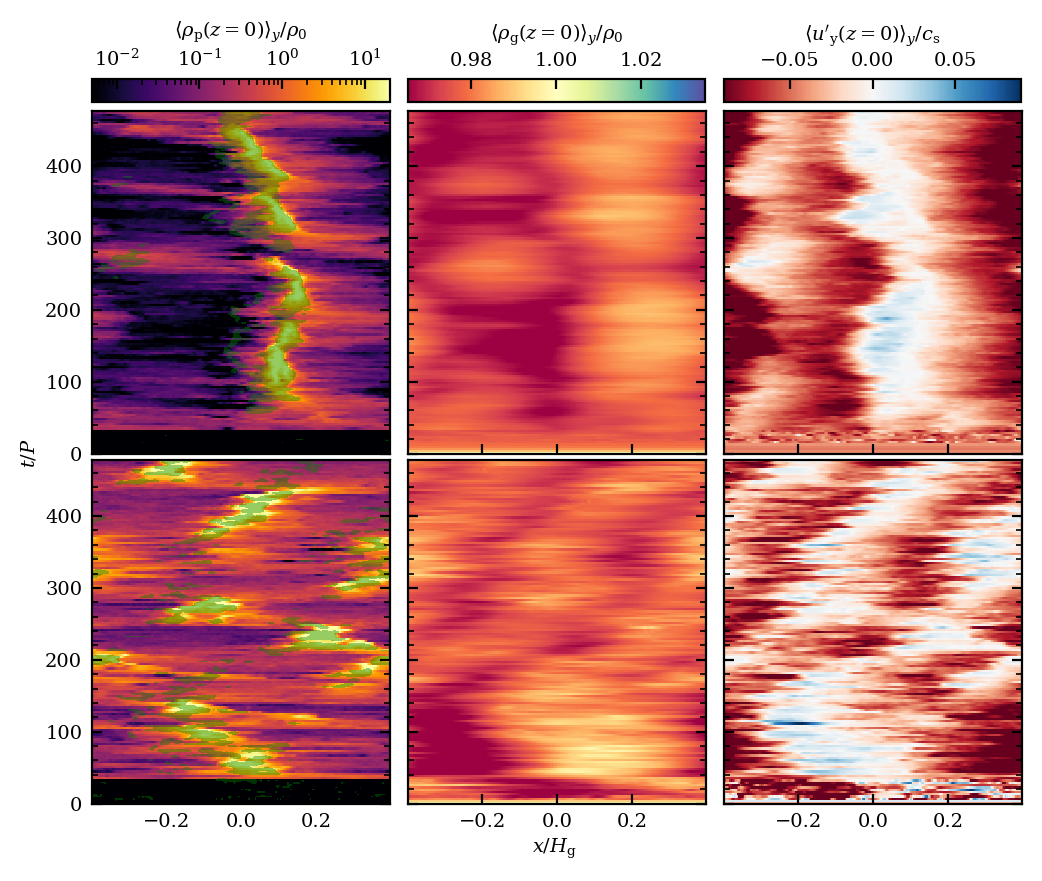}
    \caption{Time evolution of the azimuthally averaged particle density {\em (left)}, gas density (middle), and gas azimuthal velocity corrected for the background pressure gradient {\em (right)} at the midplane (vertically averaged $\pm1$ cell from $z=0$), for the strong-clumping runs with $\beta_0=6\times 10^4$, $\mathrm{St}=0.03$, $Z=0.01$ {\em (top)} and $\beta_0=6\times 10^3$, $\mathrm{St}=0.03$, $Z=0.06$ {\em (bottom)}. Particles preferentially concentrate in zonal flows (local pressure bumps), identifiable as regions where the corrected $u_y'\sim0$ in the right column. The green shaded regions in the left panels show the center of the zonal flows where the pressure gradient is close to 0 ($|u_y'|\leq0.007$)}. 
    \label{fig: tVSx 2 sim}
\end{figure*}

Zonal flows are large-scale azimuthal structures that produce density and pressure variations in the saturated MRI \citep{Johansen2009a,Bai2014ApJ...796...31B}. Local pressure variations within them translate into changes in the radial drift of particles, and zero pressure gradients in pressure maxima result in particle trapping \citep{Whipple1972}. 
The role of zonal flows in clumping is best assessed by comparing the locations of particle clumps with the locations of the zonal flows. At a pressure maximum, the gas orbits at the Keplerian velocity, so $u’_y = 0$, and the gas density reaches a local maximum. Positive pressure gradients correspond to $u’_y>0$, while negative pressure gradients correspond to $u’_y<0$. When identifying zonal flows from the gas density profile, one must keep in mind that in our model, a background pressure gradient is implemented via a radial force; consequently, any pressure maximum appears on the left, negative-$x$ side of the corresponding gas density maximum.

Figure \ref{fig: tVSx 2 sim} shows the time evolution of the azimuthally averaged particle density, gas density, and gas azimuthal velocity in the midplane for two simulations that exhibit strong clumping. Since the radial pressure gradient is added to the particles instead of the gas, we subtracted $\Pi c_{\rm s}$ from $u_y^{\prime}$ in this figure. Consistent with \citet{XuBai2022}, we find a clear correspondence between the locations of particle clumps and zonal flows. In the top panels, which show data from the simulation with $\beta_0=6\times10^4$, $\mathrm{St}=0.03$, and $Z=0.01$, a particle filament develops at $x/H_{\rm g} \simeq 0.1$ after approximately 80 orbital periods. The filament subsequently disappears and re-emerges several times, but remains at roughly the same radial location. This location lies to the left of the gas density maximum and coincides with regions where $u’_y\sim0$, as expected.

In our simulation with $\beta_0=6\times10^3$, $\mathrm{St}=0.03$, and $Z=0.06$ (bottom panels of Fig.~\ref{fig: tVSx 2 sim}), we likewise find a strong correspondence between the locations of particle filaments and zonal flows. In this case, the zonal flows - and consequently the particle filaments - are more transient and appear at different radial locations over time. We note that the presence of zonal flows does not, by itself, guarantee strong clumping, as our non-clumping runs also exhibit zonal flows.

It should be noted that at exactly $\Pi=0$ there is no radial drift, and the SI is therefore formally absent. However, the pressure bumps associated with zonal flows are dynamically active, so the zero gradient region fluctuates in position, as can be seen in Fig.~\ref{fig: tVSx 2 sim}. As a result, the SI can grow across the pressure bump. Studies of SI in idealized pressure bumps by \citet{Carrera2021} showed the growth of SI in the finite gradient regions, producing the fluctuations necessary to allow growth throughout the structure. \citet{XuBai2022} further demonstrated reduced maximum particle densities in zonal flows when SI was prevented by turning off dust drag on the gas. 

\section{Discussion}\label{sec:discussion}

\subsection{Other types of turbulence}
Turbulence in protoplanetary disks can be generated by multiple (magneto)hydrodynamical processes, with different mechanisms dominating in different regions of the disk \citep{Lesur+23}. Several turbulence-driving processes may also coexist, such as the MRI and the VSI \citep{CuiBai2022}. As demonstrated by the comparison between our simulations and those of \citetalias{Lim2024}, the forcing-scale properties of the turbulence play a crucial role in particle clumping. In other words, two different turbulence-driving mechanisms that produce the same overall turbulence strength as characterized by various turbulent parameters do not necessarily yield the same clumping threshold. This is because self-consistently driven turbulence not only introduces diffusive effects, but can also modify the disk structure, for example through the formation of pressure bumps or anticyclonic vortices. Both diffusion and disk structure strongly influence particle dynamics, and therefore the ability of the SI to concentrate particles.  

The study by \citetalias{Lim2024}, which employs forced isotropic turbulence, represents the least favorable scenario for particle clumping by the SI, since the turbulence is purely diffusive and does not generate pressure variations or anisotropies that can aid particle concentration. In contrast, in our nonideal MHD simulations including AD, clumping arises from a combination of concentration due to dust back-reaction and particle trapping within zonal flows. Such zonal flows have been reported in both local \citep{RiolsLesur2018} and global \citep{CuiBai2021} simulations of nonideal MHD with AD, as well as in simulations with ideal MHD \citep{Johansen2009a,Bai2014ApJ...796...31B}. Turbulence is significantly stronger in the ideal MHD regime, and although planetesimal formation via the SI is indeed possible under these conditions \citep{Johansen2007,Johansen2011}, such environments are expected to be found only in the innermost regions of protoplanetary disks \citep{Armitage2011}.

Turning to purely hydrodynamical processes, one mechanism that has received considerable attention in recent years is the VSI. The VSI has been shown to coexist with both the SI \citep{SchaferJohansen2022} and the MRI \citep{CuiBai2022}. Global simulations including both the SI and the VSI find that the VSI can trigger the Rossby wave instability, leading to the formation of vortices that work in concert with the SI to concentrate particles and form planetesimals \citep{HuangBai2025,HuangBai2025_jun}. 
\subsection{Convergence studies} \label{subsec:convergence}

To test the convergence of our results with resolution, we performed simulations with a resolution of 500 cells per gas scale height (marked HR in Figure \ref{fig:Zcrit}), twice our fiducial resolution. Due to the high computational cost, we end these simulations shortly after the Hill density has been reached. We confirm that doubling the spatial resolution does not lead to an increase of the required dust-to-gas ratio for strong clumping to occur, except in the case when $\beta_0=6\times 10^3$ and $\textrm{St}=0.01$. For this parameter combination, the strong clumping criterion was not reached after 400 orbital periods when using $Z=0.075$, but strong clumping was reached when increasing $Z$ to $0.1$. We further performed simulations without particle feedback, and confirmed that the time-averaged values of the turbulent parameters were similar to those on the fiducial grid (see Tab.~\ref{table: alpha}). 

To test the convergence with domain size, we perform additional simulations with a 75\% larger radial box length than the fiducial case. When using the fiducial domain size, sometimes only one or two zonal flows are present, but with the larger domain size we confirm that multiple zonal flows appear. Our results show that an increase in radial domain size does not lead to an increase of the required dust-to-gas ratio for strong clumping to occur. As in the case with high resolution, we also performed simulations without particle feedback, to confirm that the turbulent parameters do not vary too much compared to the fiducial setup (see Table \ref{table: alpha}). In this study, we do not investigate the convergence with vertical domain size. As discussed in Section \ref{subsec: results compare LI24/LI26}, modifications to $L_z$ and the vertical boundary condition have been shown to affect the clumping boundary in past studies of the pure SI. The dependence of the clumping threshold on vertical domain size remains to be investigated in future studies. 

\subsection{Limitations of the model}\label{subsec: limitations to model}
Both the MRI and SI are sensitive to numerical parameters such as domain size and spatial resolution \citep{Yang2009,Yang2012,YangJohansen2014,Li2018}. The wavelength of the fastest MRI growth is resolved by at least ten cells in our fiducial models, and we fit a minimum of six wavelengths into the vertical extent of the simulation domain \citep{Sano2004}. 
The particle scale heights in our parameter study vary between $\sim0.03H_{\rm g}$ and $\sim0.2H_{\rm g}$, as estimated from simulations with a clear pre-clumping phase, resulting in a minimum grid resolution of $\sim8H_{\rm p}^{-1}$. We adopt a particle resolution of one particle per grid cell, as done in several previous studies. Although an increased particle resolution would result in reduced Poisson noise, the number of particles per grid cell does not strongly affect clumping by the SI \citep{BaiStone2010_oct}.

Our study, which constitutes the first parameter study of the SI in the presence of self-consistently generated MRI turbulence, is limited to two magnetic field strengths and three Stokes numbers. Deriving a fit to the clumping boundary, analogous to those presented in \citetalias{Lim2024} and \citetalias{Lim2026}, would require additional simulations, which we leave for future work. Furthermore, all simulations were performed with a constant global radial pressure gradient of $\Pi=0.05$, consistent with previous studies of the SI clumping boundary. Since the pressure gradient can vary with disk radius---even in disks without local pressure variations---it would be desirable to derive a fit that explicitly accounts for its effect; this would require simulations spanning a range of pressure gradients. 

Our strong clumping criterion is based on the Hill density, which varies with location and disk mass. For example, weaker clumping may still trigger collapse farther out in the disk even if it does not in the inner disk. The clumping boundary we obtain is therefore not universal and may shift depending on disk location and disk mass.
Finally, this study assumes particles with a single $\mathrm{St}$, whereas real disks contain a distribution of particle sizes. The polydisperse SI has been explored using both linear \citep{Krapp2019,Paardekooper2021,ZhuYang2021} and nonlinear \citep{BaiStone2010,Schaffer2021,YangZhu2021} simulations, with mixed results. To date, a detailed clumping boundary for the polydisperse SI remains to be established.
 
\section{Conclusions}\label{sec:conclusion}

We present the first parameter study of the SI in 3D simulations including self-consistently generated MRI turbulence. We model a local patch of the outer protoplanetary disk, where the dominant nonideal MHD effect is ambipolar diffusion. We perform simulations with two magnetic field strengths, three Stokes numbers $\mathrm{St}=0.01$, $0.03$ and $0.1$, and varying dust-to-gas surface density ratios $Z$ in order to identify the boundary for strong clumping.
Our main findings can be summarized as follows:
\begin{itemize}
    \item In the case of modest turbulence with $\alpha_{\rm SS}\sim10^{-4}$, the clumping boundary remains close to that obtained from pure SI simulations. This demonstrates that turbulence does not necessarily hinder planetesimal formation. We find significantly subsolar $Z_{\rm crit}$ for $\mathrm{St}=0.1$, while supersolar $Z_{\rm crit}$ is required for $\mathrm{St}<0.03$. When the turbulence strength is increased to $\alpha_{\rm SS}\sim10^{-3}$, $Z_{\rm crit}$ becomes supersolar for all considered values of $\mathrm{St}$.
    \item A comparison with forced-turbulence models shows that our inferred $Z_{\rm crit}$ values are significantly lower, indicating that self-consistently generated turbulence that produces driving-scale structures can pose a smaller obstacle to planetesimal formation than isotropically forced turbulence that primarily acts to diffuse dust particles. 
    \item Clumping in our simulations occurs within zonal flows -- large-scale azimuthal density and pressure variations generated by MRI turbulence -- that act to concentrate particles and, together with dust back-reaction, generate particle clumps.
\end{itemize}


\begin{acknowledgements}
We thank P. Armitage and D. Sengupta for useful discussions.
LEJE and M-MML acknowledge support from NASA via the Emerging Worlds program under grant \#80NSSC25K7117.
ZX acknowledges support from the European Union's Horizon Europe research and innovation programme under the Marie Skłodowska-Curie grant agreement No. 101211616 (PLANDEVOL) and the Carlsberg Foundation (Semper Ardens: Advance grant FIRSTATMO).
CCY acknowledges the support from NASA via the Emerging Worlds program (\#80NSSC23K0653), the Astrophysics Theory Program (grant \#80NSSC24K0133), and the Theoretical and Computational Astrophysical Networks (grant \#80NSSC21K0497).
PH acknowledges support from the National Science Foundation of China under grant Nos. 12503070, 12233004 and 12533011.
J.L. acknowledges support from NASA under the Future Investigators in NASA Earth and Space Science and Technology grant \#80NSSC22K1322.

\end{acknowledgements}

\bibliographystyle{aa}
\bibliography{ref}

\end{document}